\documentclass[aps,prb,twocolumn,amsmat,
amssymb,amsfonts,superscriptaddress]{revtex4-1}

\usepackage{amsmath}
\usepackage{times}
\usepackage[varg]{txfonts}
\usepackage{hyperref}

\usepackage{color}
\usepackage{graphics}
\usepackage{graphicx}
\usepackage[percent]{overpic}
\usepackage{caption}
\usepackage{subcaption}
\usepackage{bm}

\newcommand{\ket}[1]{\left| #1  \right \rangle}

\newcommand{\exc}[3]{\left< #1 \vphantom{#2#3} \right|
 #2 \left| #3 \vphantom{#1#2} \right>} 

\newcommand{\avg}[1]{\langle #1 \rangle}

\newcommand{\h}[1]{{#1}^{\dagger}} 

\newcommand{\cb}[1]{\bar{#1}}

\newcommand{\up}{\uparrow}
\newcommand{\down}{\downarrow}

\newcommand{\meV}{{\rm meV}}

\newcommand{\nairo}{Na$_2$IrO$_3$}
\newcommand{\liiro}{Li$_2$IrO$_3$}

\definecolor{acolor1}{RGB}{68,34,221}
\definecolor{acolor2}{RGB}{28,181,0}
\definecolor{acolor3}{RGB}{238,34,34}

\hypersetup{colorlinks=true,linkcolor=acolor1,
citecolor=acolor1,urlcolor=acolor1} 
\begin{document}

\title{Trigonal distortion in the honeycomb iridates: Proximity of zigzag and spiral phases in \nairo{}}
\author{Jeffrey G. Rau}
\affiliation{
Department of Physics, University of Toronto, 
Toronto, Ontario M5S 1A7, Canada}
\author{Hae-Young Kee}
\email[Electronic Address: ]{hykee@physics.utoronto.ca}
\affiliation{
Department of Physics, University of Toronto, 
Toronto, Ontario M5S 1A7, Canada}
\affiliation{
Canadian Institute for Advanced Research/Quantum Materials Program, 
Toronto, Ontario MSG 1Z8, Canada}

\date{\today}
\begin{abstract}
The Kitaev honeycomb model has been suggested as a useful
starting point to describe the honeycomb iridates.
However, the zigzag ordering
seen in \nairo{} and the magnetic transition in \liiro{} have raised questions
to their connection to the Kitaev model and
to the microscopic origin of these magnetic states, given their
structural similarities.
Here we study how the magnetic phases near the Kitaev spin liquid
are affected by the inclusion of trigonal distortion of the
oxygen octahedra within a purely nearest neighbour model.
Using a combination classical analysis and exact diagonalization 
we show that near the ferromagnetic Kitaev limit a small amount of trigonal
distortion, as found in \nairo{}, stabilizes a zigzag phase.
Decreasing the distortion destabilizes the 
zigzag phase toward a spiral phase that may be relevant for \liiro{}.
Using semi-classical spin-wave calculations we show that this regime
is qualitatively consistent with experimentally known features of the
dynamical structure factor in \nairo{} and speculate on implications
for \liiro{}.
\end{abstract}
\maketitle

\section{Introduction}

The honeycomb iridates \nairo{} and \liiro{} have provided a useful playground 
to study the effects of strong spin-orbit coupling in magnetic materials \cite{
singh2010antiferromagnetic,liu2011long,singh2012relevance,ye2012direct,
lovesey2012magnetic,choi2012spin,comin2012na_,clancy2012spin,
gretarsson2013magnetic,gretarsson2013crystal,cao2013evolution}. Motivation 
originally stemmed from proposals that these compounds could be proximate to 
Kitaev's honeycomb model\cite{kitaev2006anyons}, possibly manifesting its 
exotic ground state: a gapless $Z_2$ spin liquid. Subsequent experiments
\cite{singh2010antiferromagnetic} have proven that this picture incomplete, 
with both materials exhibiting magnetic order below $\sim 15K$. In \nairo{}, 
resonant  inelastic X-ray scattering (RIXS)\cite{liu2011long,
gretarsson2013magnetic,gretarsson2013crystal} and neutron scattering 
experiments\cite{liu2011long,choi2012spin,ye2012direct} the ordering 
wave-vector of this magnetic state was determined to be the $M$ point. Further 
evidence from neutron scattering\cite{ye2012direct,choi2012spin} has shown 
unambiguously that this ordering forms alternating chains of ferromagnetically 
aligned spins -- the so-called zigzag state.

To explain the appearance of this phase, extensions to the Kitaev model have 
been proposed. The most studied adds a conventional isotropic Heisenberg 
coupling in addition to Kitaev terms giving the so-called Heisenberg-Kitaev 
(HK) model\cite{chaloupka2010kitaev,jiang2011possible,reuther2011finite,
trousselet2011effects,schaffer2012quantum,price2012critical,price2013finite,
chaloupka2013zigzag,okamoto2013global,trousselet2014hole}. If indirect oxygen mediated hopping 
is assumed to be dominant, this model produces a ferromagnetic Kitaev (FK) 
exchange. However, within the HK model 
one must lie near the antiferromagnetic Kitaev (AFK) limit 
to achieve the observed zigzag state (AFK-zigzag) -- opposite to the 
expectation from dominant oxygen mediated hopping. Other proposals consider 
using significant second and third nearest-neighbour Heisenberg couplings
\cite{kimchi2011kitaev} to stabilize a zigzag phase near the FK limit.

While theoretically appealing, the HK model is not the minimal model for the 
honeycomb iridates, even in  the idealized limit, with no trigonal or 
monoclinic distortions. As shown recently in Ref. \onlinecite{rau2014generic}, 
at nearest neighbour level an additional term must be included -- so-called 
symmetric off-diagonal exchange $\Gamma$, leading
to the HK$\Gamma$ model. Such a term is generically induced when both 
oxygen mediated and direct overlap of the $5d$ orbitals are present in the 
underlying microscopic model. This HK$\Gamma$ model has additional phases appearing 
near the FK limit, such as incommensurate spirals and further zigzag regions. 
While these results are suggestive, the trigonal compression present 
in the crystal structure of \nairo{} should be taken into account.

Recently, the effects of such trigonal distortion of the oxygen octahedra 
and the monoclinic distortions of the lattice have been considered
\cite{katukuri2014kitaev,yamaji2014honeycomb} in ab-initio treatments of 
\nairo{}. The results of these calculations suggest a dominant FK coupling, 
as well as sensitivity of the exchanges to the oxygen positions. 
While these calculations can stabilize a zigzag ground state (FK-zigzag)
\cite{katukuri2014kitaev,yamaji2014honeycomb} and qualitatively reproduce 
the susceptibility anisotropy\cite{yamaji2014honeycomb}, whether these models 
can account for the known features of the excitation spectrum seen in RIXS and 
inelastic neutron scattering (INS) experiments remains to be seen. 
In this approach, zigzag order is stabilized through the presence of additional 
anisotropic exchanges as well as further neighbour anisotropic couplings 
generated due to the inclusion of trigonal and monoclinic distortion.
Including these effects complicate the model considerably -- at the nearest 
neighbour level alone ten independent exchange constants must be considered. 
Due to the large number of phases that meet near the Kitaev limits, it is 
unclear which of these many interactions is responsible for stabilizing the 
zigzag order.

In this article we show that when trigonal distortion is included
the appearance of the zigzag state can be explained within a purely nearest 
neighbour model -- without the need to appeal to second and third neighbour couplings or 
significant monoclinic distortion.
To make the physics as transparent 
as possible we work with a model of the \nairo{} structure that allows us 
to study the effects of trigonal distortion directly. Through microscopic 
calculations and symmetry arguments we first generalize the HK$\Gamma$
model to include the effects trigonal distortion of the oxygen octahedra. 
We show that this trigonal distortion introduces an additional symmetric 
off-diagonal exchange $\Gamma'$ into the spin Hamiltonian. We then analyze 
this model near the AFK and FK limits using a classical analysis as well 
as through exact diagonalization. Two distinct zigzag regions are found: 
near the AFK limit as found in Refs. \onlinecite{chaloupka2013zigzag,
okamoto2013global} and near the FK limit with finite $\Gamma$ discussed in 
Ref. \onlinecite{rau2014generic}. The latter zigzag phase is further stabilized 
by the addition of negative $\Gamma'$. To compare with the experimental results 
we compute the dynamical structure factor using semi-classical spin-wave 
theory for each zigzag region. While the the AFK-zigzag can be made 
qualitatively consistent with the constraints from INS and RIXS experiments, 
one must tune $\Gamma$ to be small while keeping $K$ large and positive -- 
without generating an antiferromagnetic $J$. If oxygen mediated
exchange is larger or comparable to the direct $dd$ overlap this regime seems 
implausible. In the FK regime we find that a zigzag phase with
nearly gapless excitations can be stabilized with significant $\Gamma$
and small negative $\Gamma'$. In addition,  we find that when $\Gamma'$ 
is small or positive values the zigzag phase becomes unstable towards a 
multiple-$Q$ spiral with dominant wave-vector lying in the first 
Brillouin zone (BZ). We conclude that with addition of trigonal distortion 
the the FK-zigzag is qualitatively consistent with the features seen 
experimentally in \nairo{} and possibly provides a connection to the ordered 
phase of \liiro{}. 

The article is organized as follows: in Sec. \ref{sec:microscopics} 
we give an overview of the atomic physics of the Ir$^{4+}$ in the presence 
of trigonal distortion, outlining the derivation of the pseudo-spin model using 
a strong coupling expansion. Expressions for $J$, $K$, $\Gamma$ and $\Gamma'$ are presented in 
a simplified limit, with the full general case and some details presented in 
Appendix \ref{app:general}. In Sec. \ref{sec:classical} we present
simulated annealing calculations of the classical phase diagram when $\Gamma'$ 
is included, discussing the new multiple-$Q$ incommensurate spiral phases 
that appear. In Sec. \ref{sec:quantum} we focus on the FK and AFK limits 
using exact diagonalization of a $24$-site cluster, presenting phase diagrams 
as a function of $J$ and $\Gamma$ for variety of $\Gamma'$. In Sec. \ref{sec:sw} we consider 
spin-wave calculations of the properties of the FK-zigzag and AFK-zigzag 
phases found in Sec. \ref{sec:classical} and Sec. \ref{sec:quantum}, 
discussing the connection to reported experimental results for the dynamical structure 
factor from INS and RIXS experiments. In Sec. \ref{sec:discussion} we discuss 
the implications for \nairo{} as 
well as the applicability of these results to \liiro{}. 

\section{Microscopics}
\label{sec:microscopics}
To derive an effective model that captures the essential physics of \nairo{} 
and \liiro{} we build our description using an idealized version of the 
crystal structure. We start with an ideal honeycomb lattice surrounded 
by edge-shared oxygen octahedra, with all monoclinic and trigonal distortions 
removed. Moving beyond this, we include the effects of trigonal distortions 
of the oxygen octahedra. We must first identify the relevant degrees of 
freedom at the ${\rm Ir}$ site.
\subsection{Local physics}
\label{sec:local}
Consider a single ${\rm Ir}{\rm O}_6$ octahedron in the absence of interactions; the crystal 
field provided by the ${\rm O}^{2-}$ splits the $5d$ orbitals of the ${\rm Ir}^{4+}$ ion into an $e_g$ 
doublet and $t_{2g}$ triplet. Since the electronic configuration is $5d^5$ we have a 
single hole in the $t_{2g}$ and unoccupied $e_g$ levels. Since the energy scale of the 
octahedral splitting is on the order of a few ${\rm eV}$, the $e_g$ levels can be safely 
neglected. The remaining $t_{2g}$ levels form a pseudo-vector with the angular 
momentum of the $d$ electrons projected into the $t_{2g}$ subspace is given by $-\vec{L}$ 
where $\vec{L}$ are $l=1$ angular momentum matrices. The single-particle parts of 
the atomic Hamiltonian have the form
\begin{equation}
  - \lambda \vec{L}\cdot \vec{S}
  + \Delta \left(\hat{n} \cdot \vec{L}\right)^2,
\end{equation}
where $\hat{n}$ is a unit vector along the $[111]$ direction.  The sign of $\Delta$ 
distinguishes between trigonal compression ($\Delta>0$) and trigonal expansion 
($\Delta<0$) of the oxygen octahedra. Since $\hat{n}\cdot(\vec{L} + \vec{S})$ is conserved, we rotate the 
spin and effective orbital angular momentum so that $\hat{z}$ is along $\hat{n}$. In this 
basis the Hamiltonian is easily diagonalized (see Ref. 
\onlinecite{khaliullin2005orbital} for details) giving three doublets
\begin{subequations}
\begin{eqnarray}
  \label{eq:doublets}
  \ket{1,\pm} &=& \cos\theta \ket{\frac{1}{2},\pm \frac{1}{2}}_{\hat{n}} \pm \sin\theta \ket{\frac{3}{2}, \pm \frac{1}{2}}_{\hat{n}}\\
  \ket{2,\pm} &=& \mp \sin\theta \ket{\frac{1}{2},\pm \frac{1}{2}}_{\hat{n}} + \cos\theta \ket{\frac{3}{2}, \pm \frac{1}{2}}_{\hat{n}}\\
  \ket{3,\pm} &=& \ket{\frac{3}{2},\pm \frac{3}{2}}_{\hat{n}}
\end{eqnarray}
\end{subequations}
where $\ket{j,m}_{\hat{n}}$ are the effective $j=1/2$ and $j=3/2$ states quantized with $\hat{z}$ 
along the $[111]$ direction. The angle $\theta$ parametrizes the relative strength of 
trigonal distortion and spin-orbit coupling 
\begin{equation}
  \tan(2\theta) = \frac{4 \sqrt{2} \Delta}{2 \Delta + 9 \lambda}.
\end{equation}
Note that for small trigonal distortion $\theta \sim 2\sqrt{2} \Delta/9\lambda$. The energies of these 
states are given by
\begin{subequations}
\begin{eqnarray}
  E_1 &=& \frac{1}{2}\left(\frac{\lambda}{2} + {\Delta}\right) +\sqrt{\frac{1}{4}\left(\frac{\Delta}{3}+\frac{3\lambda}{2}\right)^2 +\frac{2\Delta^2}{9}},\\
  E_2 &=& \frac{1}{2}\left(\frac{\lambda}{2} + {\Delta}\right) -\sqrt{\frac{1}{4}\left(\frac{\Delta}{3}+\frac{3\lambda}{2}\right)^2 +\frac{2\Delta^2}{9}},\\
  E_3 &=& {\Delta} - \frac{\lambda}{2}.
\end{eqnarray}
\end{subequations}
In the absence of spin-orbit coupling (near $\theta = \tan^{-1} (2\sqrt{2})/2$) this term splits 
the $t_{2g}$ levels in an $a_{1g}$ singlet and $e_g$ doublet, with the energy difference 
given by $\Delta$. When both trigonal distortion and spin-orbit coupling are 
included the $t_{2g}$ levels split into three Kramers doublets with the degeneracy 
fully lifted. In this case the relevant degrees of freedom are in the 
highest-lying doublet which is half-filled. When trigonal distortion is 
zero (near $\theta=0$), the three $t_{2g}$ orbitals form an effective $l=1$ pseudo-vector, 
with spin-orbit coupling splitting these into a $j=1/2$ doublet and $j=3/2$
quartet. The five $d$ electrons fill the $j=3/2$ states completely, leaving a 
single electron in the $j=1/2$ doublet. This $j=1/2$ doublet adiabatically 
connect to the $\ket{1,\pm} \equiv \ket{\pm}$ doublet for all values of $\Delta/\lambda$ and thus represents 
our low energy degrees of freedom.

We note that the full distortion (both monoclinic and trigonal) 
of the oxygen octahedra in \nairo{} lowers the symmetry
at the $\rm Ir$ site to $2/m$: a single $C_2$ 
rotation and reflection. In \nairo{}, the trigonal distortion is seen clearly
if one looks at the difference between in the oxygen-oxygen distances for
bonds that lie in the $[111]$ plane and those that do not\cite{choi2012spin}. 
While the in-plane bonds are separated by $\sim 3.01 - 3.03 \AA$, the other bonds vary from
$2.65 - 2.97 \AA$. In \liiro{} the site symmetry
is reduced to a single two-fold rotation, 
a single $C_2$ rotation\cite{o2008structure}.  The
the difference between the bonds in the $[111]$ plane and 
the others is also less pronounced, although there is still large variations
between the different bonds due to monoclinic distortion. This is consistent
with estimates of the change of the trigonal distortion in these compounds from
magnetic susceptibility measurements\cite{cao2013evolution}.
Generically, inclusion all of these distortions gives three Kramers doublets,
as in the case with just trigonal distortion.
For simplicity we will ignore such further small corrections to the $\ket{\pm}$ doublet 
from these monoclinic distortions, as our goal is to isolate and study the 
effects of trigonal distortion.

\begin{figure}[tp]
  \includegraphics[width=0.8\columnwidth]{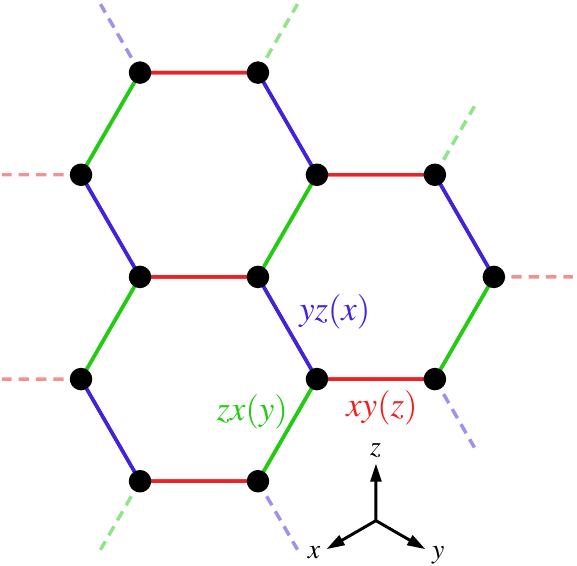}
\caption{
\label{fig:exchanges}
Notation for the Kitaev and bond-dependent exchanges.
We have denoted the $yz(x)$ bonds blue,
the $zx(y)$ bonds green and the $xy(z)$ bonds red. The
choice of $x$,$y$ and $z$ axes is also shown. }
\end{figure}
\subsection{Tight-binding model}
\label{sec:tight}
With the local degrees of freedom identified, we now consider hopping 
processes between ${\rm Ir}$ atoms. This contribution to the Hamiltonian has the form
\begin{equation}
H_{\rm kinetic}=  \sum_{\avg{ij}} \sum_{\alpha\beta} \h{d}_{i\alpha} t^{\alpha\beta}_{ij} d_{j\beta},
\end{equation}
where $\h{d}_{i\alpha} = (\h{d}_{i\alpha \up}\ \h{d}_{i\alpha \down})$ and $d_{i\alpha}$ are the creation and annihilation operators for 
the $t_{2g}$ state $\alpha$ at site $i$. To clarify what is possible, we first look at how 
the symmetry of the  model constrains the hopping terms. Consider a nearest 
neighbour ${\rm Ir}$-${\rm Ir}$ bond of type $xy(z)$ as shown in Fig. \ref{fig:exchanges}. The 
relevant processes that contribute to these hoppings will involve mainly the 
two ${\rm Ir}^{4+}$ ions and their respective ${\rm O}^{2-}$ octahedral cages. The symmetries of 
this complex then constraint the hopping matrix. When the octahedra are ideal, 
these symmetries include inversion through the bond center, time-reversal as 
well as $C_2$ rotations along the $[1\cb{1}0]$ and $[001]$ axes. For example, 
time-reversal and inversion force the matrix to be both real and symmetric, 
while the presence of the $C_2$ axis along $[1\cb{1}0]$ requires $t^{yz,xy} = t^{zx,xy}$ and 
$t^{zx,zx}=t^{yz,yz}$. If the all symmetries are used, the hopping matrix on this bond 
is then constrained to the form
\begin{equation}
  t_{xy(z)} =   \left( \begin{tabular}{ccc}
      $t_1$ & $t_2$ & $0$ \\
      $t_2$ & $t_1$ & $0$ \\
      $0$ & $0$ & $t_3$ 
      \end{tabular}
    \right),  
\end{equation}
where the basis is ordered $yz$, $zx$ and $xy$. The presence of the $C_2$ about 
$[001]$ is what prevents any mixing between the $xy$ orbital and the $zx$ and $yz$ 
orbital, forcing $t^{yz,xy} = t^{zx,xy} = 0$. If we consider the symmetries of the true 
space group $C2/m$ which are present, then this $C_2$ about $[001]$ is lost. This 
results in an additional allowed term, giving the more general form
\begin{equation}
  t_{xy(z)} =   \left( \begin{tabular}{ccc}
      $t_1$ & $t_2$ & $t_4$ \\
      $t_2$ & $t_1$ & $t_4$ \\
      $t_4$ & $t_4$ & $t_3$ 
      \end{tabular}
    \right).  
\end{equation}
When monoclinic distortion is neglected, the kinetic terms for the $yz(x)$ and 
$zx(y)$ bonds can be found using the rotational symmetry of the lattice from 
those of the $xy(z)$ bond. Rotating the $t_{2g}$ orbitals we then arrive at the 
hopping matrices
\begin{equation}
  t_{yz(x)} =   \left( \begin{tabular}{ccc}
      $t_3$ & $t_4$ & $t_4$ \\
      $t_4$ & $t_1$ & $t_2$ \\
      $t_4$ & $t_2$ & $t_1$ 
      \end{tabular}
    \right), \ \ \ 
  t_{zx(y)} =   \left( \begin{tabular}{ccc}
      $t_1$ & $t_4$ & $t_2$ \\
      $t_4$ & $t_3$ & $t_4$ \\
      $t_2$ & $t_4$ & $t_1$ 
      \end{tabular}
    \right).
\end{equation}
When the distortions are included the $yz(x)$ and $zx(y)$ must be analyzed 
separately, with hopping parameters distinct from the $xy(z)$ bond. Further, the
$yz(x)$ and $zx(y)$ bonds do not possess the $C_2$ axis and thus are only required 
to be real and symmetric. However, in this case the $yz(x)$ and $zx(y)$ are not 
independent as the $C_2$ axis about the $xy(z)$ bond relates the hopping matrices.
Putting these all together, the kinetic Hamiltonian can be then written in the compact form
\begin{eqnarray}
  \label{eq:kinetic}
\sum_{\avg{ij}\in \alpha\beta(\gamma)} &&\left[
    t_1\left(\h{d}_{i\alpha} d_{j\alpha}+\h{d}_{i\beta} d_{j\beta}\right)+
    t_2\left(\h{d}_{i\alpha} d_{j\beta}+\h{d}_{i\beta} d_{j\alpha}\right)+
    t_3 \h{d}_{i\gamma} d_{j\gamma}+\right. \nonumber\\
&&\left.
    t_4 \left(\h{d}_{i\gamma} d_{j\alpha}+\h{d}_{i\gamma} d_{j\beta}+
    \h{d}_{i\alpha} d_{j\gamma}+\h{d}_{i\beta} d_{j\gamma}\right)
    \right],
\end{eqnarray}
where we sum over the $yz(x)$, $zx(y)$ and $xy(z)$ links as indicated in 
Fig. \ref{fig:exchanges}, mapping the directions to orbitals as $x \rightarrow yz$, $y \rightarrow zx$ 
and $z \rightarrow xy$. 

There are several processes that generate such kinetic terms for the $\rm Ir$ $t_{2g}$ 
orbitals. Two important mechanisms are direct $dd$ overlap and oxygen mediated 
hopping. Considering again the $xy(z)$ bond, each type of hopping can be written 
using a Slater-Koster scheme\cite{slater1954simplified}, expressing the 
amplitudes in terms of the $dd$ and $pd$ parameters. We begin with the fully 
idealized structure.  Given the large $t_{pd\pi}$ overlap,
we have taken the oxygen-mediated parts to be dominant. This 
contributes an $zx$-$yz$ inter-orbital hopping of order $t^2_{pd\pi}/\Delta_{pd}$  where $\Delta_{pd}$ is 
the ${\rm Ir}$-${\rm O}$ charge gap. When projected into the $j=1/2$ subspace this 
inter-orbital term vanishes, so we must consider further contributions. 
The simplest to include is direct overlap of the $5d$ orbitals which induces 
both intra- and inter-orbital terms. Putting these contributions together 
we have
\begin{subequations}
\begin{eqnarray}
  t_1 = t^{yz,yz} &=&  \frac{t_{dd\pi}+t_{dd\delta}}{2}, \\
  t_2 = t^{yz,zx} &=&  \frac{t^2_{pd\pi}}{\Delta_{pd}}+ \frac{t_{dd\pi}-t_{dd\delta}}{2},  \\
  t_3 = t^{xy,xy} &=&  \frac{3t_{dd\sigma}+t_{dd\delta}}{4}.
\end{eqnarray}
\end{subequations}
Further contributions are possible to each of $t_1$, $t_2$ and $t_3$. 
For example, hopping mediated through the central ${\rm Na}$ 
introduces a contribution to $t_2$ of order $t_{sd\sigma}^2/\Delta_{sd}$ where $t_{sd\sigma}$ is the overlap 
between the ${\rm Na}$ $3s$ and the ${\rm Ir}$ $5d$ orbitals and  $\Delta_{sd}$ being the 
${\rm Na}$-${\rm Ir}$ charge gap. 
The addition of trigonal compression of the oxygen octahedra allows for $t_4$ to 
become non-zero. This occurs since the oxygens tilt out of the $xy$ plane of 
the bond, breaking the $C_2$ symmetry about $[001]$. The
inclusion of trigonal distortion thus induces two distinct effects: the change in 
the local atomic states and through the generation of these additional kinetic terms.

\subsection{Strong-coupling limit}
\label{sec:strong}
The relevant degree of freedom in the strong-coupling limit is a pseudo-spin
$\vec{S}_i$ at each site originating from the half-filled doublet. In the most 
idealized case, this doublet effectively has $j=1/2$ and so the spin operators 
transform like a pseudo-vector. This remains true even as trigonal distortion 
is introduced. Considering the symmetry of the $xy(z)$ bond discussed in the 
previous section, the allowed exchanges are constrained in  manner identical 
to that of the $t_{2g}$ hoppings, as they transform in the same way under symmetry. 
We thus see that without trigonal distortion the allowed exchanges for a $xy(z)$ 
from site $i$ to $j$ are of three types: Heisenberg exchange $J$, Kitaev exchange $K$ 
and symmetric off-diagonal exchange $\Gamma$. Making this relationship explicit, the 
exchange $t_1$ is analogous to $J$, the exchange $t_3$ to $J+K$ and the $t_2$ to the 
exchange $\Gamma$. Extending this to $yz(x)$ and $zx(y)$ bonds we have the spin Hamiltonian
\begin{equation}
\label{hkg-model}
\sum_{\avg{ij} \in \alpha\beta(\gamma)} \left[ 
  J \vec{S}_i \cdot \vec{S}_j + 
  K S^\gamma_i S^\gamma_j +
  \Gamma (S^\alpha_iS^\beta_j+S^\beta_iS^\alpha_j)
\right].
\end{equation}
When trigonal distortion is included then we have an addition term;
we will call this contribution
$\Gamma'$ and it takes the form
\begin{equation}
\label{hkgg-model}
  \Gamma' \sum_{\avg{ij}\in \alpha\beta(\gamma)} \left[
S^\alpha_i S^\gamma_j + S^\gamma_iS^\alpha_j+
S^\beta_i S^\gamma_j + S^\gamma_iS^\beta_j
    \right].
\end{equation}
In the language of symmetry this is analogous to the $t_4$ contribution to the 
hopping Hamiltonian. In the absence of monoclinic distortion this is most 
general nearest-neighbour spin model allowed for these doublets. \footnote{As in the case of hoppings the
introduction of monoclinic distortion causes the $xy(z)$ bond to be inequivalent
to the $yz(x)$ and $zx(y)$ bonds. This allows for an independent set of these exchanges
on these bonds, and two different $\Gamma'$ couplings.}

To see how these exchanges arise from the underlying microscopic theory we 
carry about a strong-coupling expansion. There are several limits depending on 
the order in which the energy scales are taken to be large. Two limits under 
which this is tractable analytically are $U,J_H \gg \lambda, \Delta \gg t$ and $U,\lambda \gg J_H \gg \Delta \gg t$. 
We will consider the former case as it is most commonly used in the literature
and illustrates how each contribution appears in the exchanges. We stress that 
the results do not depend strongly on which limit is used. To begin, we assume 
an atomic Hamiltonian of Kanamori form\cite{sugano1970multiplets}:
\begin{equation}
  \label{eq:atomic}
  H_0 = \sum_i\left[ \frac{U-3 J_H}{2} (N_i-5)^2 - 2J_H S_i^2 -\frac{J_H}{2} L_i^2\right],
\end{equation}
where $N_i$, $S_i$, and $L_i$ are the total number, spin, and (effective) orbital 
angular momentum operators at site $i$, $U$ is the Coulomb interaction, and $J_H$ is 
Hund's coupling. Since the spin-orbit coupling and trigonal distortion then 
dominate the kinetic terms, the resulting spin-orbital model can be projected 
into the subspace of the $\ket{1,\pm}$ doublet\footnote{We choose a basis for this
doublet subspace so that magnetic moment operator is proportional to 
effective spin operators $\vec{S}_i$ up a a diagonal $g$-factor matrix.} 
shown in Eq. \ref{eq:doublets}.
As in Section \ref{sec:local} we will work with the quantization axis along 
the $[111]$ direction, so the states shown in Eq. \ref{eq:doublets} can be used 
directly. Note that this requires the hoppings to be rotated into this basis as 
well. Treating the kinetic terms as a perturbation yields the Hamiltonian
given in Eqs. \ref{hkg-model}-\ref{hkgg-model}. 
Due to the complexity of expressions we leave a full presentation of these expressions for Appendix
\ref{app:general} and consider some simple limiting regimes. We will treat the 
trigonal distortion as small, expanding the expressions to order $\theta$. We first 
consider the case where the oxygen mediated and $dd\sigma$ overlap are the dominant 
kinetic processes with $t_1,t_4 \sim 0$ and $t_2,t_3 \neq 0$.
\begin{eqnarray}
  \label{eq:exchanges3}
  J &=& 
\frac{4}{27}\left[
    \frac{2 t_3^2+4\sqrt{2}\theta t_2 t_3}{U-J_H}
    -\frac{
      12\sqrt{2}\theta t_2 t_3
    }{U-3J_H}
    +
    \frac{
      t_3^2+4\sqrt{2}\theta t_2 t_3
    }{U+2 J_H}
    \right], \nonumber
    \\
  K &=& -\frac{8 J_H }{9}\left[
    \frac{ 3t_2^2 -t_3^2 -
      2 \sqrt{2} \theta t_2 t_3  }
    {(U-3J_H)(U-J_H)} 
\right],\nonumber
    \\
  \Gamma &=& -\frac{8 J_H}{9} \left[
    \frac{2t_2t_3
+\sqrt{2} \theta \left(t_2^2+t_3^2\right)}
{(U-3J_H)(U-J_H)}
\right],\nonumber
\\
  \Gamma' &=& -\frac{8 J_H}{9} \left[
\frac{\sqrt{2} \theta\left((t_2+t_3)^2+4  t_2^2\right) }
{2(U-3J_H)(U-J_H)}\right].
\end{eqnarray}
The leading terms of these expressions have the same form as seen in Ref. 
\onlinecite{rau2014generic}, with additional contributions of order $\theta$. 
 Considering the microscopic origins, we expect 
$t_2 \sim t^2_{dp\pi}/\Delta_{pd} > 0$  and $t_3 \sim t_{dd\sigma} < 0$ with
$t_2> |t_3|$. This leads to $K < 0$, $\Gamma>0$ with $J$ and $\Gamma'$ being
subleading in $t_2$. The contribution to $\Gamma'$ is directly proportional to $\theta$ 
with $\Gamma' < 0$ appearing for trigonal compression and $\Gamma' > 0$ appearing 
for trigonal expansion. As $J$ is subleading, its sign will depend
on the detailed strengths of $t_1$, $t_3$ and the other contributions
that are given in in Appendix \ref{app:general}. As most of our subsequent 
results are insensitive to the sign of $J$, we will not try to pin down its
value more precisely.
We note that additional contributions to $\Gamma'$ 
 could change the sign of $\Gamma'$, see
for example the full expressions in Appendix \ref{app:general}.
\section{Classical Phase diagram}
\begin{figure*}[tbp]
  \begin{subfigure}{0.32\textwidth}
  \includegraphics[width=\textwidth]{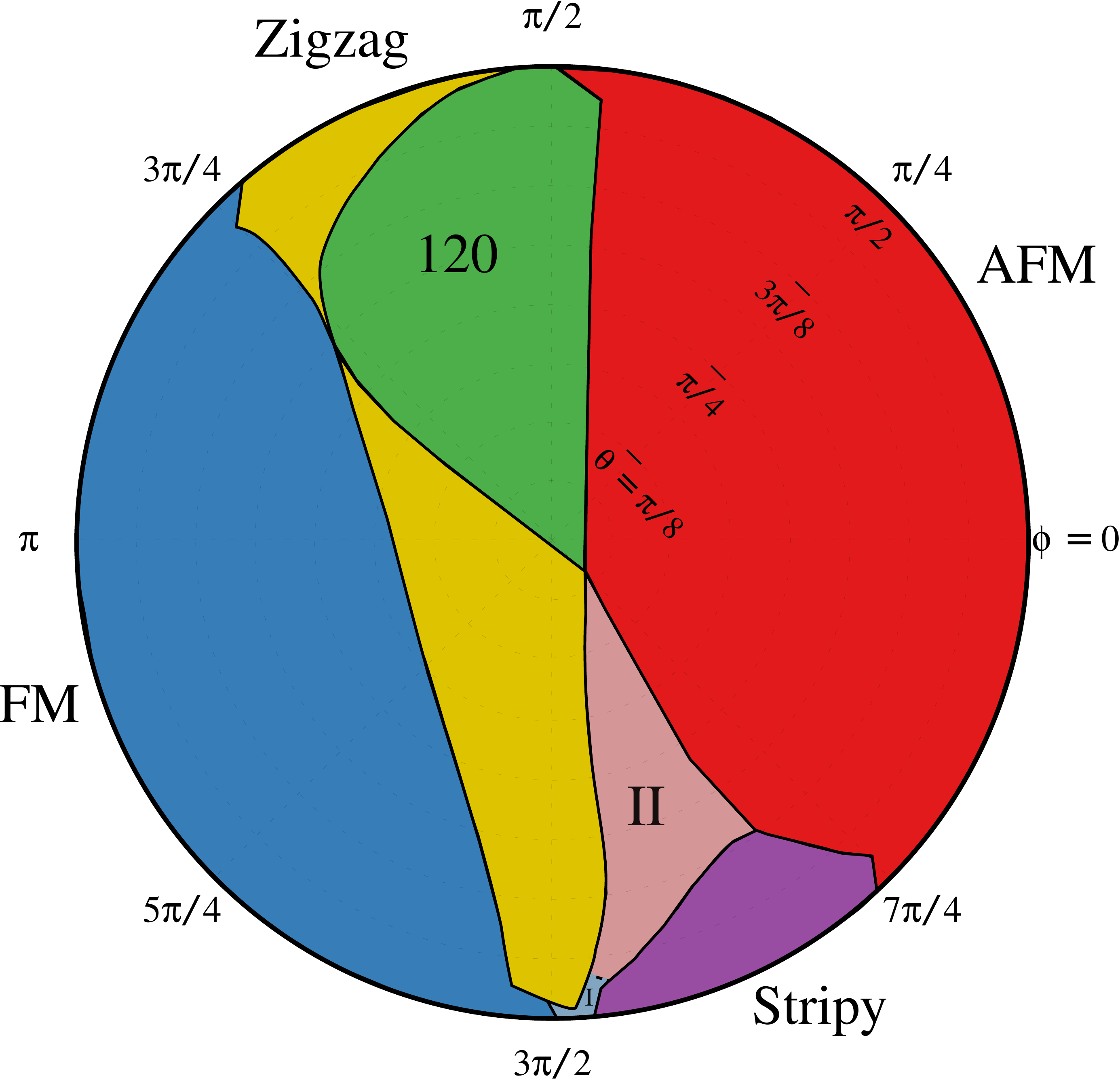}
  \caption{$\Gamma' = -0.05$\label{fig:m0p05}}
  \end{subfigure}
  \begin{subfigure}{0.32\textwidth}
  \includegraphics[width=\textwidth]{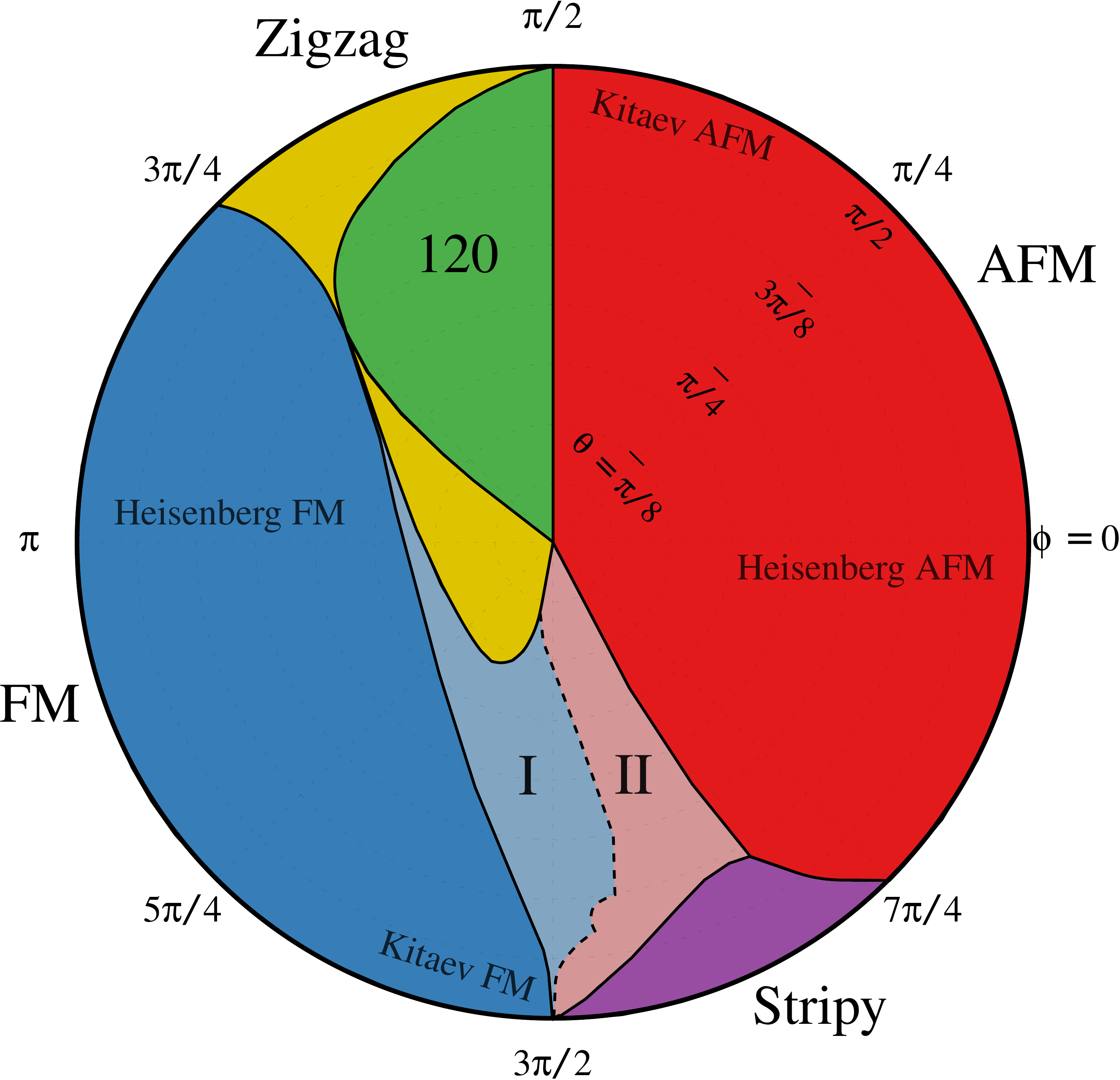}
  \caption{$\Gamma' = 0$\label{fig:0p00}}
  \end{subfigure}
  \begin{subfigure}{0.32\textwidth}
  \includegraphics[width=\textwidth]{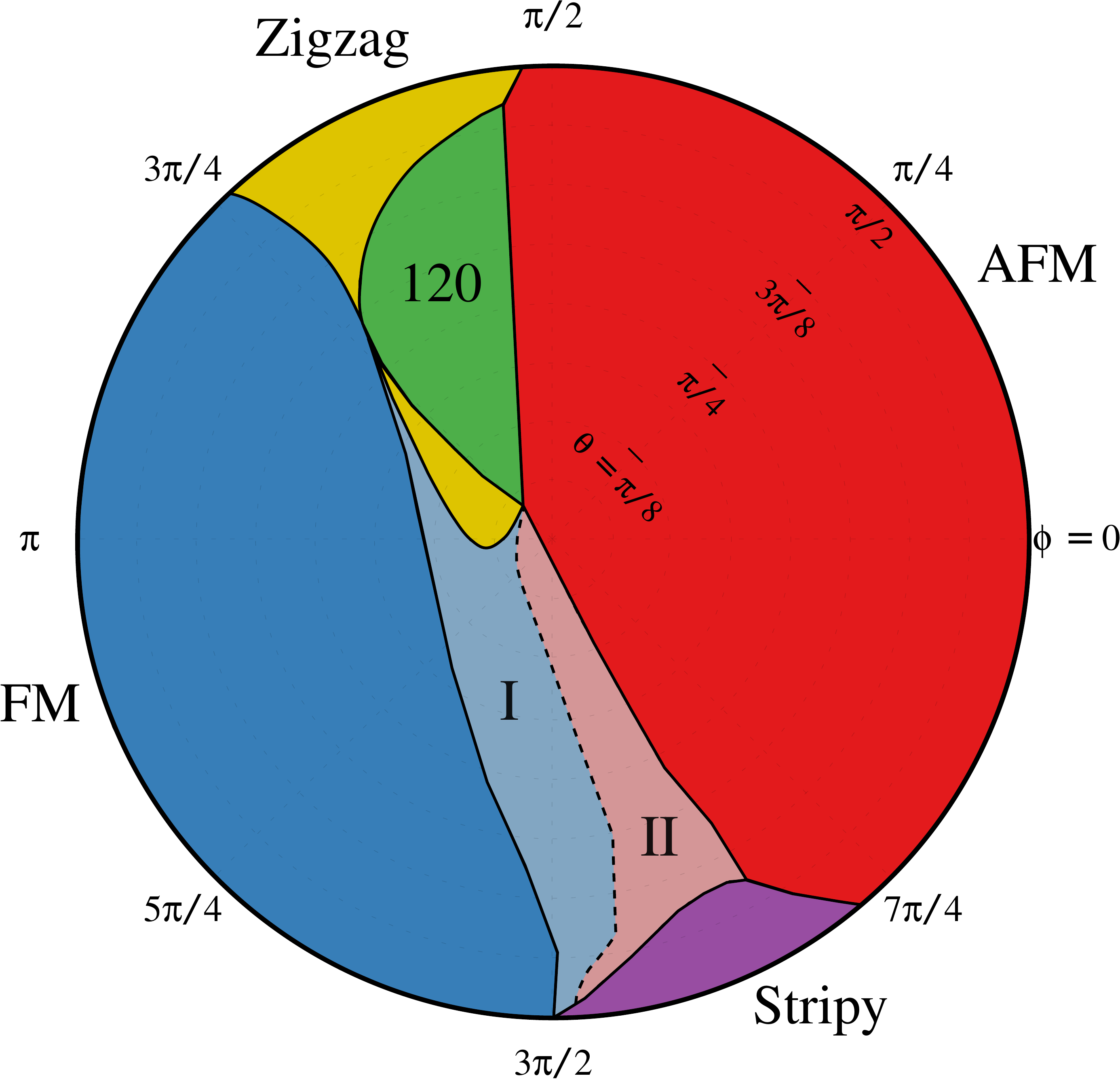}
  \caption{$\Gamma' = +0.05$\label{fig:0p05}}
  \end{subfigure}
  \begin{subfigure}{0.32\textwidth}
  \includegraphics[width=\textwidth]{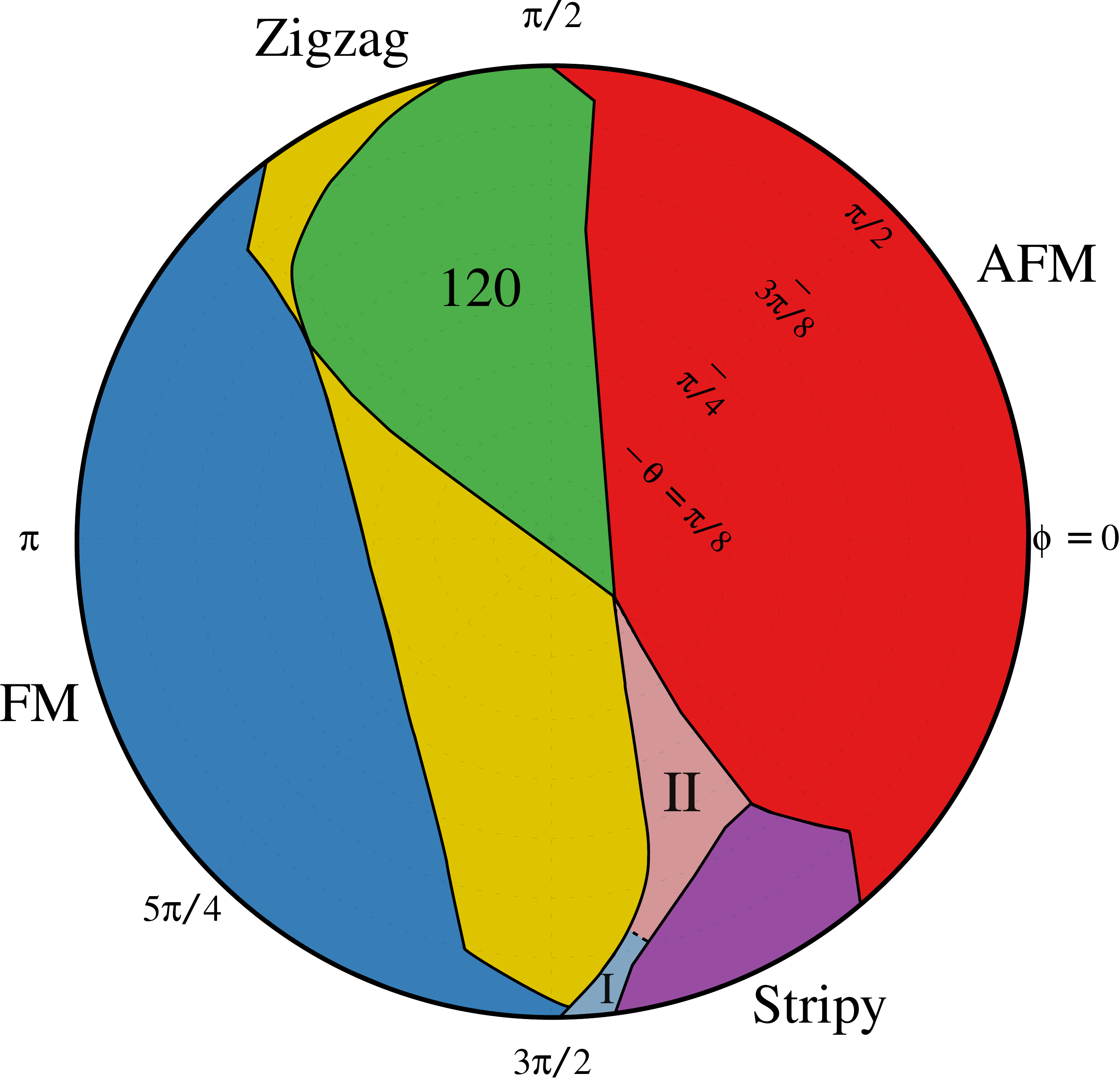}
  \caption{$\Gamma' = -0.10$\label{fig:m0p10}}
  \end{subfigure}
  \begin{subfigure}{0.32\textwidth}
  \includegraphics[width=\textwidth]{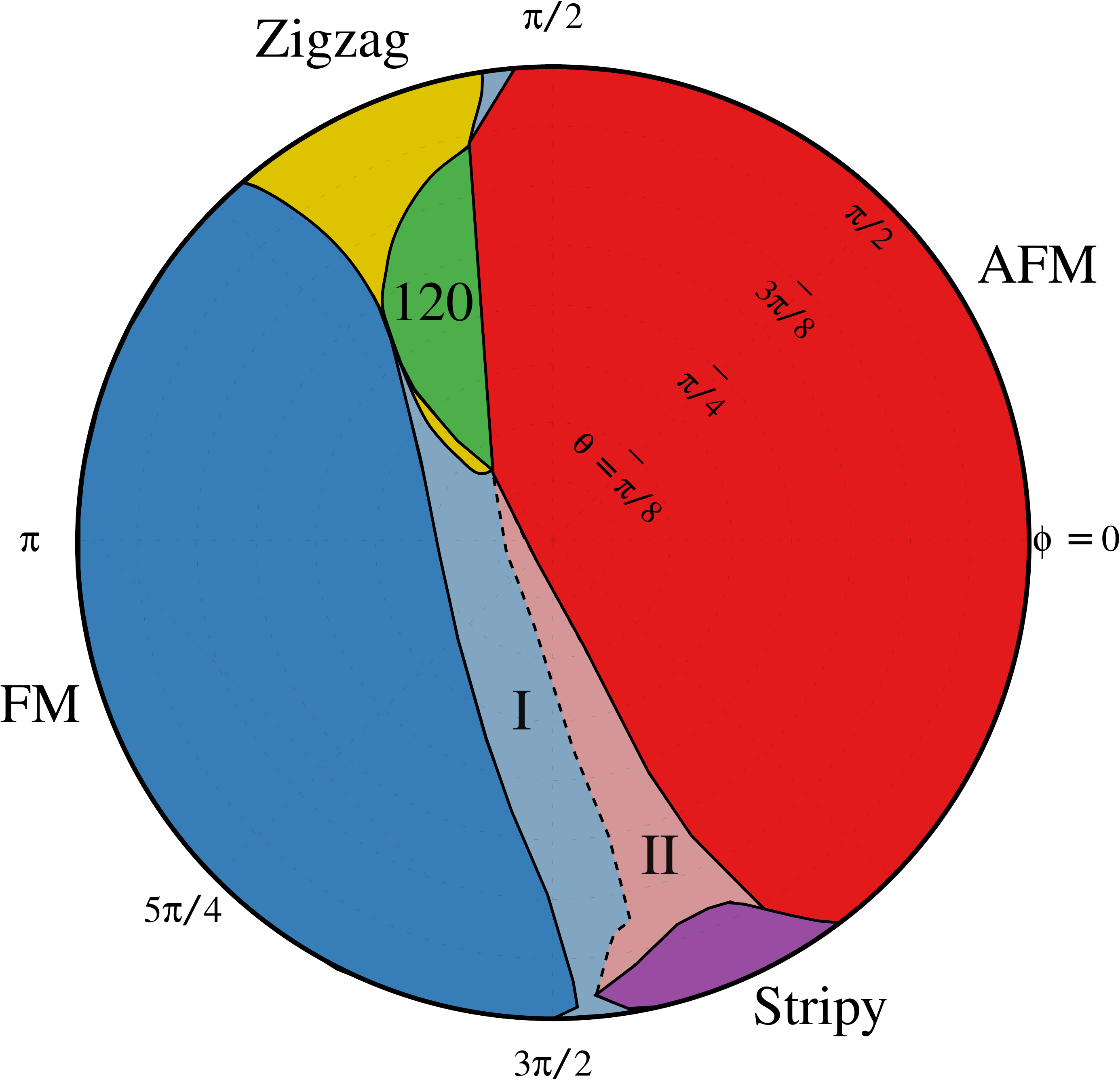}
  \caption{$\Gamma' = +0.10$\label{fig:0p10}}
  \end{subfigure}  
 \begin{subfigure}[b]{\textwidth} 
    \begin{subfigure}[t]{0.18\textwidth} 
      \includegraphics[width=\textwidth]{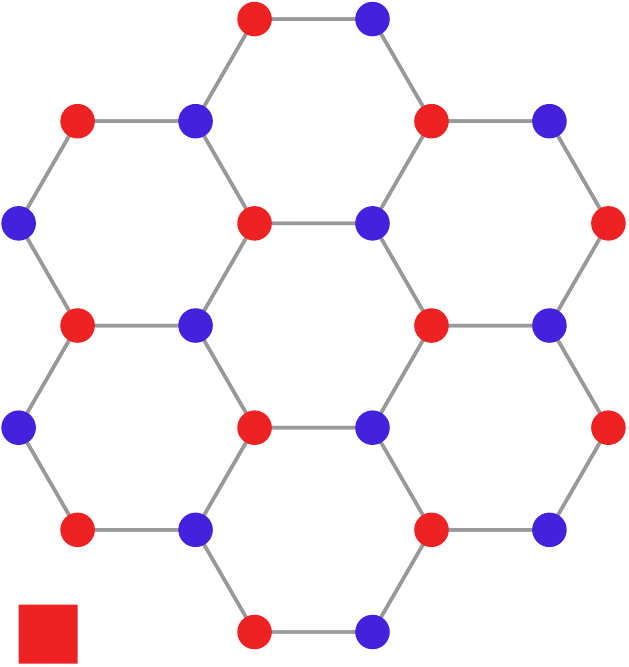}
      \caption{\label{fig:cl-afm}AFM (red)}
    \end{subfigure}
    \begin{subfigure}[t]{0.18\textwidth} 
      \includegraphics[width=\textwidth]{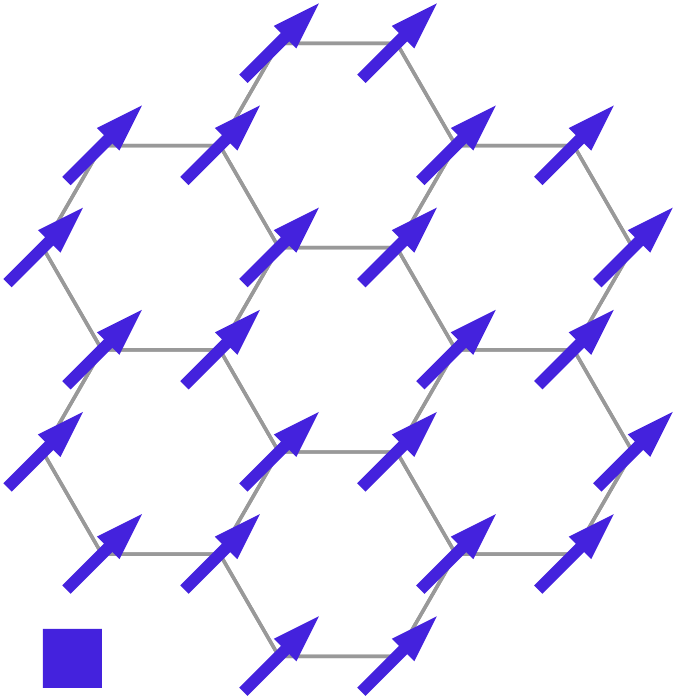}
      \caption{\label{fig:cl-fm}FM (blue)}
    \end{subfigure}
    \begin{subfigure}[t]{0.18\textwidth} 
      \includegraphics[width=\textwidth]{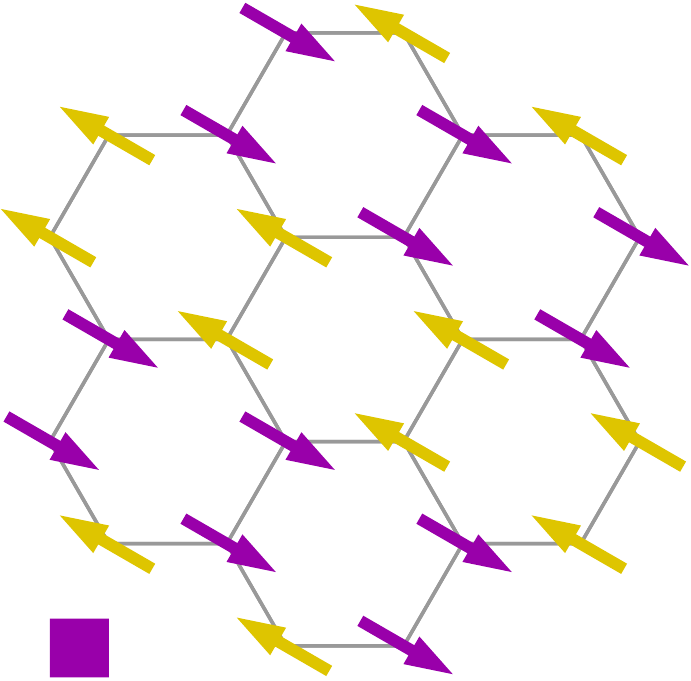}
      \caption{\label{fig:cl-stripy}Stripy (purple)}
    \end{subfigure}
    \begin{subfigure}[t]{0.18\textwidth} 
      \includegraphics[width=\textwidth]{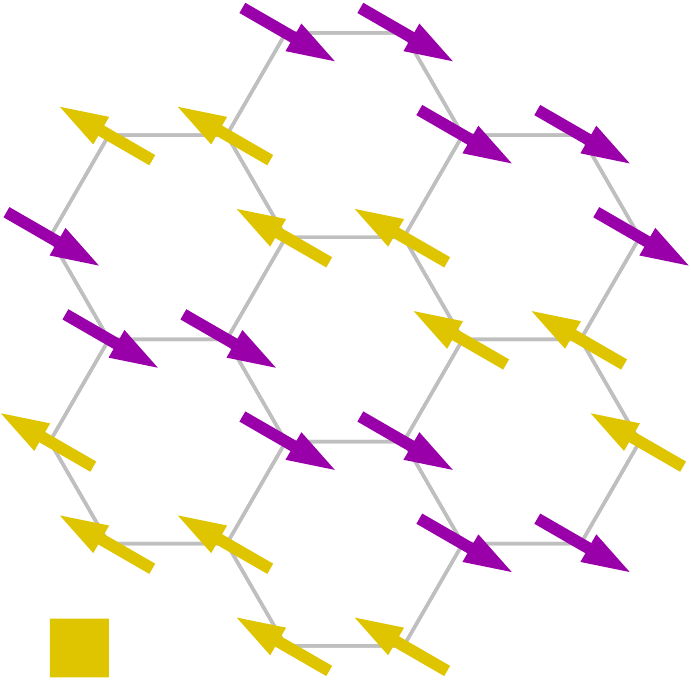}
      \caption{\label{fig:cl-zigzag}Zigzag (gold)}
    \end{subfigure}
    \begin{subfigure}[t]{0.18\textwidth} 
      \includegraphics[width=\textwidth]{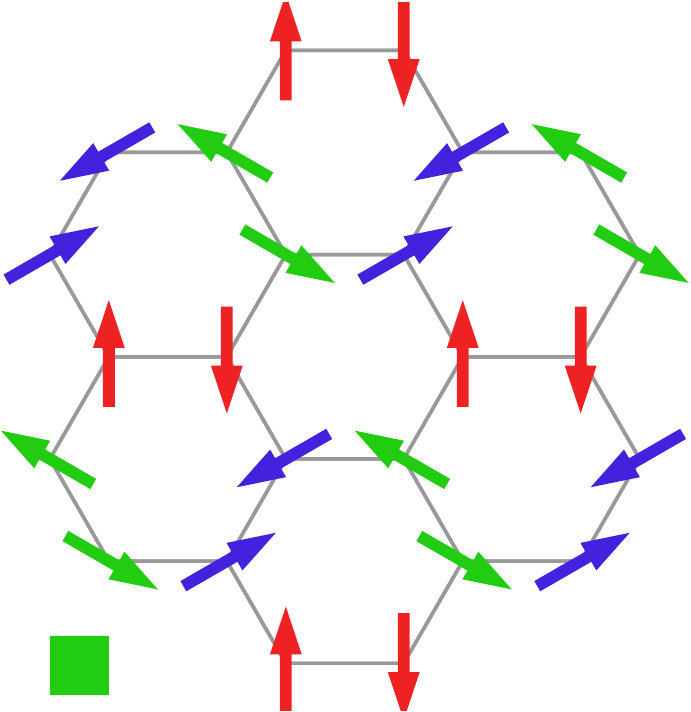}
      \caption{\label{fig:cl-120}120$^\circ$ (green)}
    \end{subfigure}
  \end{subfigure}  
\caption{
\label{fig:classical-phases} 
(a-e) Classical phase diagrams obtained through simulated annealing (see 
main text for details) for a variety of values of $\Gamma'$.
(f-g) Representative ground state spin configurations in each classical
commensurate phase, with the corresponding colour in the phase diagrams of
(a-e) given in brackets.}
\end{figure*}
\label{sec:classical}
To gain some understanding of this model  we first map out the classical phase 
diagram for arbitrary $J$, $K$ and $\Gamma$ with small $\Gamma'$. To do this we use simulated 
annealing on finite clusters with $2\cdot 12^2$ and $2\cdot 24^2$ sites to relax to the 
classical ground state using a single-spin Metropolis updating procedure.
For each point in parameter space we perform $2\cdot10^6$ sweeps with $10$ different 
starting points, taking the state with lowest energy as the classical ground 
state. The results of the two clusters are in qualitative agreement throughout 
the phase diagram.

The full phase diagrams for several values of $\Gamma'$ are shown in Fig. 
\ref{fig:classical-phases}. We have normalized the energy scale so that $J^2+K^2+\Gamma^2=1$ and parametrized
\begin{equation}
  J = \sin{\theta}\cos{\phi}, \hspace{0.5cm} K = \sin{\theta} \sin{\phi}, \hspace{0.5cm} \Gamma = \cos{\theta},
\end{equation}
allowing $\Gamma'$ to vary freely. As in the single-$Q$ analysis of Ref. 
\onlinecite{rau2014generic} we find five commensurate phases: the ferromagnet 
(FM), antiferromagnet (AFM), stripy, zigzag and 120$^\circ$ are shown in Figs. 
\ref{fig:cl-afm}-\ref{fig:cl-120}. Beyond these single-$Q$ phases, there are 
two families of multiple-$Q$ incommensurate spiral phases that appear at finite 
$\Gamma$ and $\Gamma'$, we have labeled these phases I and II. We have roughly classified 
these phases through the wave-vector of the largest components of the static 
structure factor: if it lies in the first Brillouin zone then it is in phase I, 
while if it lies outside then it is labeled II. Each of these phases has 
additional peaks and the position of dominant wave-vector seems to vary continuously 
throughout each phase. By this we mean that as the cluster size is increased 
the number of distinct dominant wave-vectors in each phase increases; the regions
of a given wave-vector become smaller and the behaviour becomes more like a smooth
gradient. We note that for small $\Gamma'$ there is a large region
in phase I with dominant wave-vector $2M/3$ in our finite size calculations. 
Within the clusters used it is unclear whether this
parameter regime is truly commensurate or incommensurate in the thermodynamic
limit, as the wave-vector could be simply changing too slowly for our finite
size calculations to resolve.  In the case with larger negative $\Gamma' = -0.10$ this dominant 
wave-vector appears to tune continuously through from $\vec{Q}=0$ (FM) to the 
maximal $\vec{Q}$ wave-vector (AFM). While the zigzag phase does not appear adjacent 
to the FK limit for $\Gamma'=0$, the nearby phase I is highly sensitive to the 
presence of $\Gamma'$. For even a small negative $\Gamma'$ we see in Fig. \ref{fig:m0p05} that 
phase I becomes unstable to the zigzag order. This persists for larger negative $\Gamma'$,
with a zigzag phase close to the FK limit, as seen in Fig \ref{fig:m0p10}.
With positive $\Gamma'$ the opposite occurs; the zigzag phase is suppressed giving way to 
larger I, II, FM and AFM phases, as seen in Figs \ref{fig:0p05} and \ref{fig:0p10}. 
We note that the wave-vector of phases I and II are dependent on the value of
$\Gamma'$.
Given the 
weakness of phases I and II, as well as the enhancement of the zigzag phase by 
quantum fluctuations\cite{rau2014generic} the relevance of these phases to the 
full quantum mechanical model at $\Gamma'=0$ remains unclear. We note that at least 
classically both the AFK-zigzag and FK-zigzag phase border incommensurate spiral
phases.

\section{Quantum Phase diagram}
\label{sec:quantum}
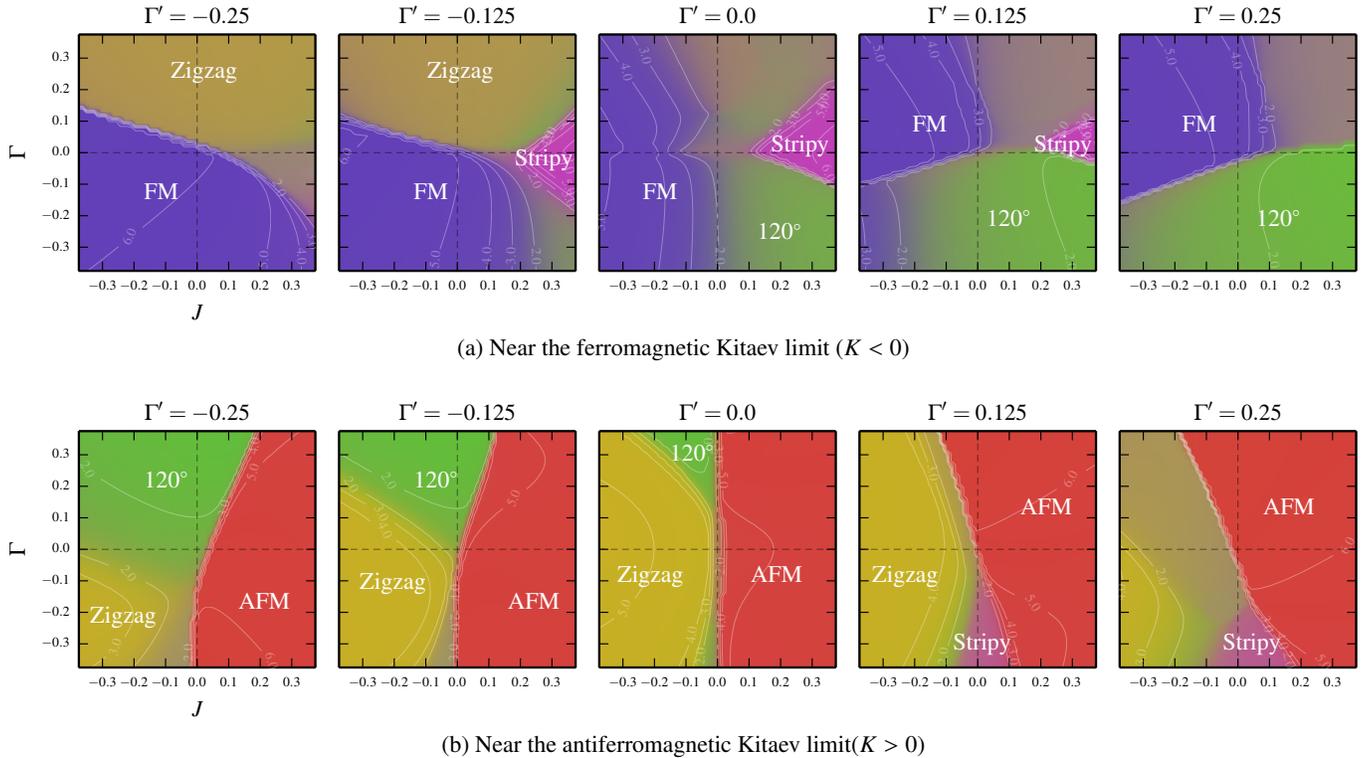
\begin{figure*}[t]
  \begin{subfigure}[b]{\textwidth}
    \begin{subfigure}[b]{\textwidth}
      \begin{overpic}[width=\textwidth]
        {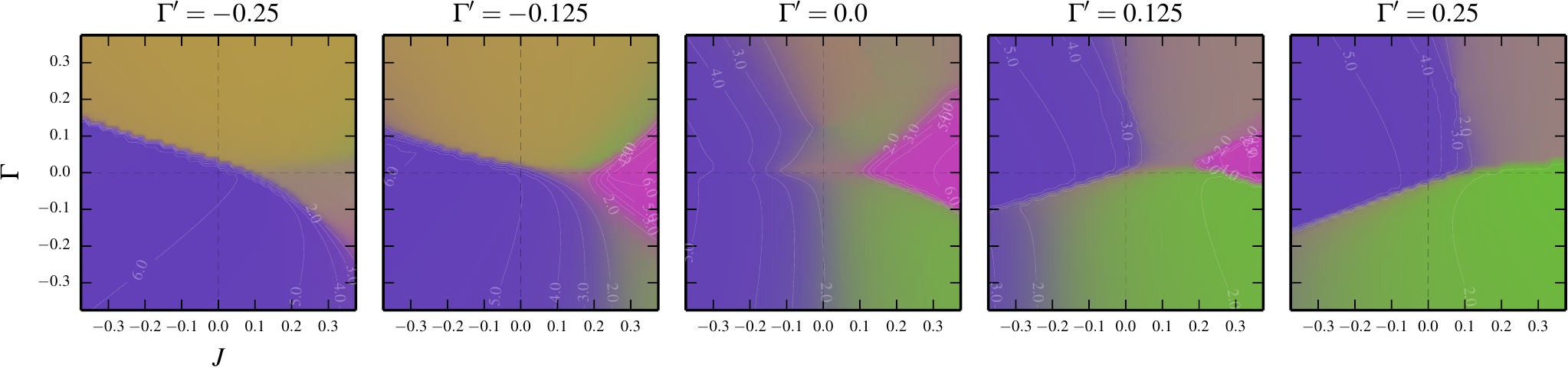}
        \put(12,18){\textcolor{white}{Zigzag}}
        \put(10,9){\textcolor{white}{FM}}

        \put(31,18){\textcolor{white}{Zigzag}}
        \put(30,9){\textcolor{white}{FM}}
        \put(37.5,11.5){\textcolor{white}{Stripy}}

        \put(47,9){\textcolor{white}{FM}}
        \put(56.5,12.5){\textcolor{white}{Stripy}}
        \put(55.5,6){\textcolor{white}{120$^\circ$}}

        \put(67,14){\textcolor{white}{FM}}
        \put(76,12.5){\textcolor{white}{Stripy}}
        \put(72.5,7){\textcolor{white}{120$^\circ$}}

        \put(87,14){\textcolor{white}{FM}}
        \put(92.5,7){\textcolor{white}{120$^\circ$}}
        \end{overpic}
      \caption{Near the ferromagnetic Kitaev limit ($K<0$)}
    \end{subfigure}
  \vspace{0.1cm}
  \end{subfigure}
  \begin{subfigure}[b]{\textwidth}
    \begin{subfigure}[b]{\textwidth}
      \begin{overpic}[width=\textwidth]
        {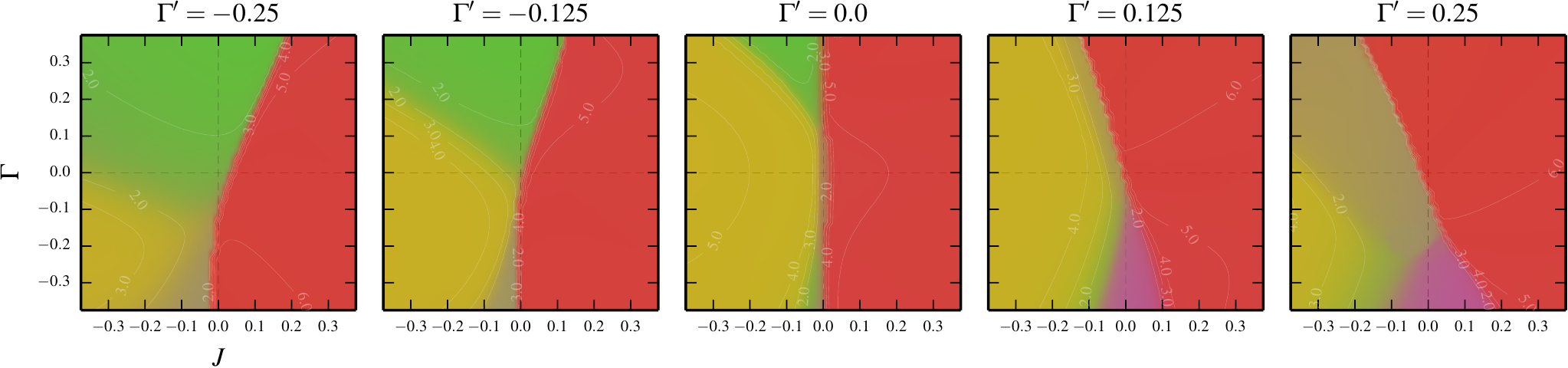}
      \put(10,17){\textcolor{white}{120$^\circ$}}
      \put(17,8){\textcolor{white}{AFM}}
      \put(6,7){\textcolor{white}{Zigzag}}

      \put(30,17){\textcolor{white}{120$^\circ$}}
      \put(37,8){\textcolor{white}{AFM}}
      \put(26,9.5){\textcolor{white}{Zigzag}}

      \put(49,19){\textcolor{white}{120$^\circ$}}
      \put(55,10){\textcolor{white}{AFM}}
      \put(45.1,10){\textcolor{white}{Zigzag}}

      \put(75,15){\textcolor{white}{AFM}}
      \put(64,10){\textcolor{white}{Zigzag}}
      \put(70,5){\textcolor{white}{Stripy}}

      \put(93,15){\textcolor{white}{AFM}}
      \put(90,5){\textcolor{white}{Stripy}}
      \end{overpic}
      \caption{Near the antiferromagnetic Kitaev limit($K>0$)}
    \end{subfigure}
  \end{subfigure}
  \caption{
\label{fig:phases}
Phase diagrams from exact-diagonalization of the 24-site cluster.
Results are shown near the (a) ferromagnetic ($K<0$) and (b) antiferromagnetic ($K>0$)
Kitaev limits as a function of $J$, $\Gamma$ and $\Gamma'$ with the
energy scale fixed so $|K|=1$. Colours identify the phases: FM (blue), AFM (red),
zigzag (gold), stripy (magenta), 120$^\circ$ (green).
The colour corresponds to values of the static structure
factor $S_Q$ in the original or rotated basis. 
Contours of constant $S_Q$ are shown for the dominant wave-vector
in each region.
}
\end{figure*}

We now compare this classical analysis to exact diagonalization study of a
$24$-site cluster. This cluster has been used in previous studies of the HK
\cite{chaloupka2010kitaev,chaloupka2013zigzag,okamoto2013global}
and HK$\Gamma$\cite{rau2014generic} model as it is compatible with most the classical 
orders of the model as well as the four-sublattice duality transformation the relates the FM and AFM phases to the stripy and zigzag. The phases were identified 
by examining the spin-spin correlation functions $\avg{S^{\alpha}_i S^{\beta}_j}$, primarily through the 
static structure factor
\begin{equation}
  \label{eq:structurefactor}
S(\vec{Q}) = \frac{1}{N} \sum_{ij} e^{i \vec{Q}\cdot(\vec{r}_i-\vec{r}_j)} \avg{\vec{S}_i\cdot\vec{S}_j},
\end{equation}
in both the original basis and after applying the four-sublattice rotation. 
We note that this four-sublattice rotation is not a duality transformation
 when $\Gamma$ or $\Gamma'$ is included, but still provides a useful indicator given the 
structure of the well-understood stripy and zigzag phases in the HK limit.
Motivated by the large diversity of phases that meet at the AFK and FK limits 
in the classical calculations, as well as the ab-initio results of 
Refs. \onlinecite{katukuri2014kitaev,yamaji2014honeycomb} we will focus on the 
FK and AFK limits in our exact diagonalization. To this end we fix the energy 
scale so that $K=\pm 1$, leaving three parameters $J$, $\Gamma$ and $\Gamma'$. We show these phase
diagrams for slices of constant $\Gamma'$ near $\Gamma'=0$, with $J$ and $\Gamma$ varying, as seen 
in Fig. \ref{fig:phases}.

Due to the qualitative similarity between the classical and quantum results,
we can directly identify the FK-zigzag and AFK-zigzag in these results. The 
zigzag phase near the AFK limit is connected to the zigzag phase seen in the HK 
model as studied in Refs. \onlinecite{chaloupka2013zigzag,okamoto2013global}. 
Appearing only when $J<0$, this phase is stable to finite $\Gamma$, but is eventually 
suppressed at large enough $\Gamma'$. The FK-zigzag only appears when $\Gamma$ is finite as 
noted in Refs. \onlinecite{katukuri2014kitaev,rau2014generic} but is quite weak 
in these exact diagonalization calculations. We see here that the addition of 
negative $\Gamma'$ stabilizes this phase; significantly enlarging the zigzag region. 
To gain further insight into these phases, we look to the structure of
excitations above the zigzag ground states.

\section{Spin waves}
\label{sec:sw}
The most detailed information on the magnetic state found in \nairo{} are based 
on RIXS and INS data\cite{gretarsson2013magnetic,choi2012spin}. The high energy spin-wave branch seen in 
RIXS\cite{gretarsson2013magnetic} with an energy scale of $\sim 30-40{\rm meV}$ points 
towards exchanges on the order of hundreds of $K$. This makes direct extraction of 
information from the susceptibility difficult as current experiments do not 
probe the high-temperature regime. Further, the energies seen in RIXS are quite 
broad, leaving uncertainty in the dispersion of this mode. The INS data
\cite{choi2012spin} is more informative, showing two well-defined features: 
magnetic excitations down to $2 {\rm meV}$ and a drop off in scattering at low energy 
with a concave edge in $\omega-|Q|$ space. The presence of these low energy 
excitations bounds the spin-wave gap, constraining any anisotropic terms in the 
spin Hamiltonian. While quite limited due the lack of directional dependence, 
the INS remains the only experimental input into the low energy magnetic 
excitations of \nairo{}.

To connect our model with these scattering experiments we will use leading 
order semi-classical spin-wave theory. Within this approximation we compute
the inelastic structure factor in the zigzag phases found in the previous 
sections. Given the  strong similarity between the classical and exact diagonalization 
phase diagrams we expect the spin-wave results to be qualitatively correct. At 
leading order in $1/S$ we express the spin operators using the 
Holstein-Primakoff representation
\begin{equation}
  \vec{S}_r \sim \left(S-\h{a}_r a_r\right)\hat{z}_r +
  \sqrt{\frac{S}{2}}\left[ \left(\hat{x}_r-i\hat{y}_r\right) \h{a}_r+
    \left(\hat{x}_r+i\hat{y}_r\right) {a}_r\right],
\end{equation}
where $a_r$ and $\h{a}_r$ are the Holstein-Primakoff bosons
and $(\hat{x}_r\ \hat{y}_r \hat{z}_r)$ define a local frame at site $r$ with $\hat{z}_r$ being the local magnetic 
ordering direction. The dynamic spin structure factor (at zero temperature) 
is proportional to
\begin{equation}
  S^{\mu\nu}(\vec{Q},\omega) \propto \sum_{n\neq 0} \delta(\omega-E_n) \exc{0}{S^\mu_Q}{n}\exc{n}{S^\nu_{-Q}}{0},
\end{equation}
where $S_Q^\mu$ is the Fourier transform of the spin operator $S^\mu_r$ and $\mu = x,y,z$.
We will be interested in the inelastic neutron scattering cross section
which can be expressed in terms of the structure factor via
\begin{equation}
  I(\vec{Q},\omega) = \frac{d^2\sigma}{d\Omega d\omega} \propto \sum_{\mu\nu} \left(1-\frac{Q_\mu Q_\nu}{Q^2}\right)S^{\mu\nu}(\vec{Q},\omega).
\end{equation}
For simplicity we have not included anisotropic $g$ factors in these 
expressions. 
We present these results showing a plot of the spin wave spectrum 
around the path $X$-$\Gamma$-$Y$-$\Gamma'$-$M$-$\Gamma$\cite{chaloupka2013zigzag}.
The colours indicate the magnitude of $I(\vec{Q},\omega)$ after convolving the structure factor with a gaussian of finite width to emulate finite experimental 
resolution. 
\section{Discussion}
\begin{figure*}[t] 
\begin{subfigure}{1\columnwidth}
  \includegraphics[width=\textwidth]{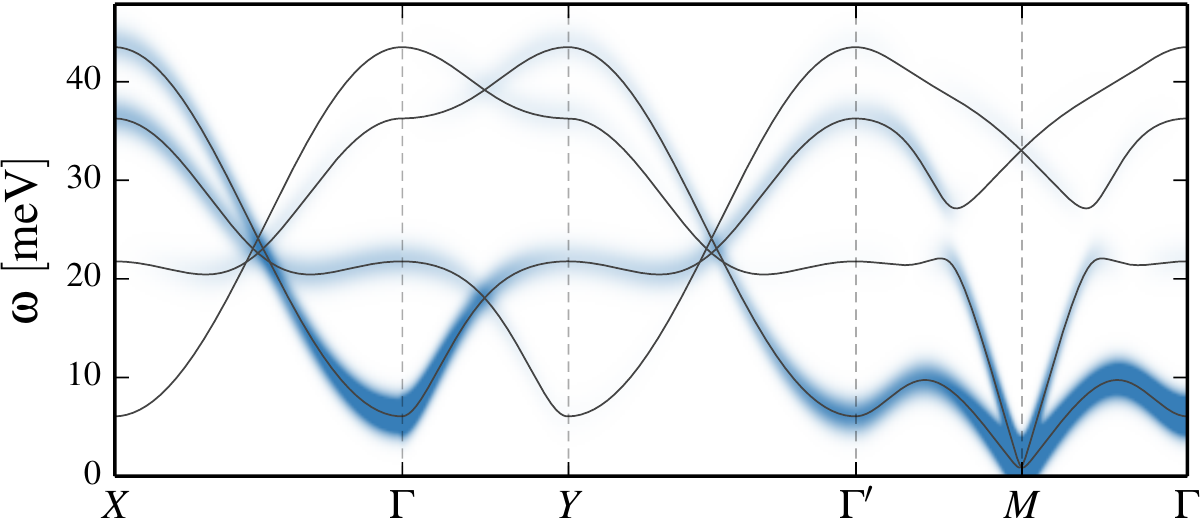}
  \caption{\centering Spin-wave spectrum (FK-zigzag)\label{fig:sw-fk}}
\end{subfigure}\hspace{0.25cm}
\begin{subfigure}{1\columnwidth}
  \includegraphics[width=\columnwidth]{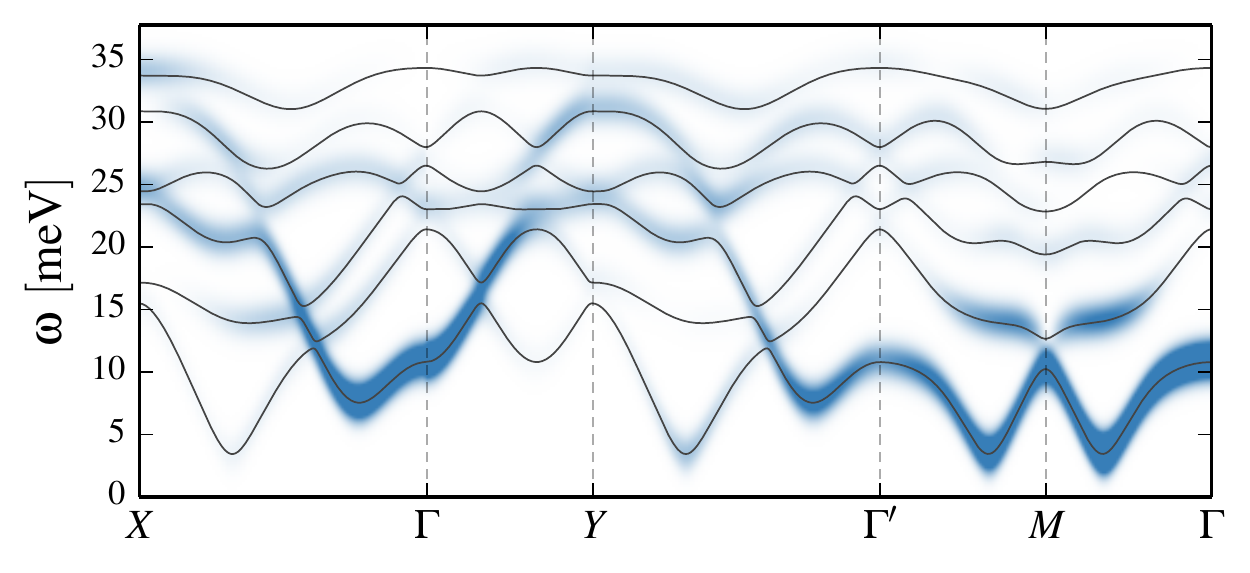}
  \caption{\centering Spin-wave spectrum ($2M/3$)\label{fig:sw-inc}}
\end{subfigure}
\caption{
Spin waves for 
(a) the FK-zigzag with 
$J=-8\meV$, $K=-25\meV$, $\Gamma=20\meV$ and $\Gamma'=-2\meV$. 
and for (b) the $2M/3$ state of phase I with
$J=-3\meV$, $K=-25\meV$, $\Gamma=20\meV$ and $\Gamma'=0\meV$. 
 Spin-wave spectrum with the inelastic cross-section $I(\vec{Q},\omega)$ shown convolved 
with a gaussian to aid visualization.}
\end{figure*}

\label{sec:discussion}
First, let us discuss the dependence of the $M$ point spin-wave gap on the 
parameters $J$,$K$, $\Gamma$ and $\Gamma'$ in each of the zigzag phases. In the AFK limit, 
earlier studies\cite{chaloupka2013zigzag} have pointed out the accidental 
SO(3) degeneracy of the classical ground state manifold in the HK limit.
This pseudo-symmetry manifests in the semi-classical calculations through 
the appearance of gapless excitations near the $M$ point -- even though spin 
rotation symmetry is strongly broken by spin-orbit effects. The addition 
of $\Gamma$ and $\Gamma'$ affect these pseudo-Goldstone modes differently. 
Moving away 
from the HK limit via $\Gamma$ immediately gaps out the $M$ point, with the gap 
equal to $\sim |\Gamma|$. We see then we can bound $|\Gamma|$ to be smaller than $\sim 1-2 {\rm meV}$ 
due to the low energy cutoff to the INS data. Curiously, adding $\Gamma'$ to the 
AFK-zigzag state does not gap out pseudo-Goldstone modes; the SO(3)
degeneracy remains unbroken. Ignoring then the microscopic route to the AFK-zigzag 
regime, the INS data can be made qualitatively consistent with this phase 
so long as $\Gamma$ is small, irrespective of the value of $\Gamma'$. 
Further, to get spin-waves that match the scales seen in the RIXS 
experiments one needs a large value for $K$.
For a representative point in the AFK-zigzag
phase, we choose $J=-10\meV$, $K=40\meV$, $\Gamma = 1\meV$ and $\Gamma'=5 \meV$.
The spin-wave spectrum for these parameter values is qualitatively similar to
that reported for the AFK-zigzag in the HK-model\cite{chaloupka2013zigzag}, 
expect for a small gap opened by finite $\Gamma$ and the splitting of some
accidental degeneracies in the spin-wave bands.

The case of the FK limit is more interesting. We first note that within the 
classical and semi-classical calculations a meta-stable zigzag phase appears 
over a wide region of parameter space that connects directly to the FK point. 
This meta-stable state is close in energy to phase I and accounts for its 
fragility under the addition of a small negative $\Gamma'$. Given the enhancement 
of the zigzag order seen in the exact diagonalization calculations we will 
discuss this zigzag phase on equal footing with the stable zigzag phase 
seen when $\Gamma'$ is finite and sufficiently negative.  Start from the FK limit 
with $J=\Gamma=\Gamma'=0$. As we increase $\Gamma$ the zigzag phase immediately opens a 
gap, with a narrow band of low energy excitations; a remnant of the flat 
band present at the FK point. As $\Gamma$ is increased further the gap reaches 
a maximum then begin decreasing again -- finally becoming gapless at the $M$ 
point at $\Gamma = 4|K|/5$. This is independent of $J$, one finds a line of gapless 
points in the meta-stable zigzag phase. At finite $\Gamma'$one can still tune to 
this regime (independent of $J$), but the zigzag phase is still not the true 
ground state. To render such a zigzag state stable at the classical level 
one must add a negative $\Gamma'$. This has the effect of opening a gap $\sim |\Gamma'|/2$ 
in the spin-waves  at the $M$ point (the proportionality constant varies 
weakly as a function of $J$). Due to the closeness in energy of phase I and 
the zigzag, the required $\Gamma'$ is small and can be consistent with the INS 
bound so long as $|\Gamma'|$ is less than $\sim 1-2 {\rm meV}$. We have shown an example of 
the spin-wave spectrum of such an FK-zigzag phase in Fig. \ref{fig:sw-fk}.
The weight of each spin-wave branch is indicated as it appears in the dynamical 
structure factor by the intensity of the color (convolved with a gaussian to aid 
visualization).  Experimentally\cite{choi2012spin}, 
one observes a region with little scattering at low momenta, bounded by a curve that
appears to terminate near $|\vec{Q}| = |M|$. The low-energy, high-weight branch that
runs from $M$ to $\Gamma$ in Fig. \ref{fig:sw-fk} qualitatively reproduces such behaviour. We note that this zigzag phase persists for positive $J$, keeping these
low-energy features, but differing in some high-energy details. 

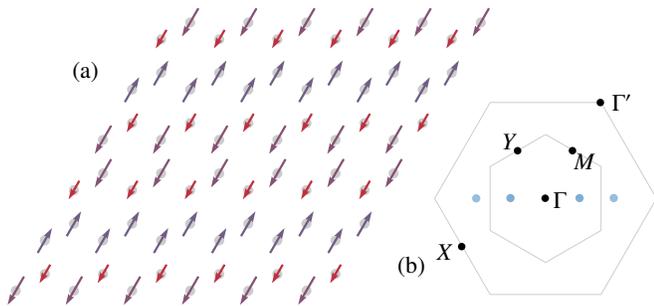
\begin{figure}[tp]
  \begin{overpic}[width=\columnwidth]
    {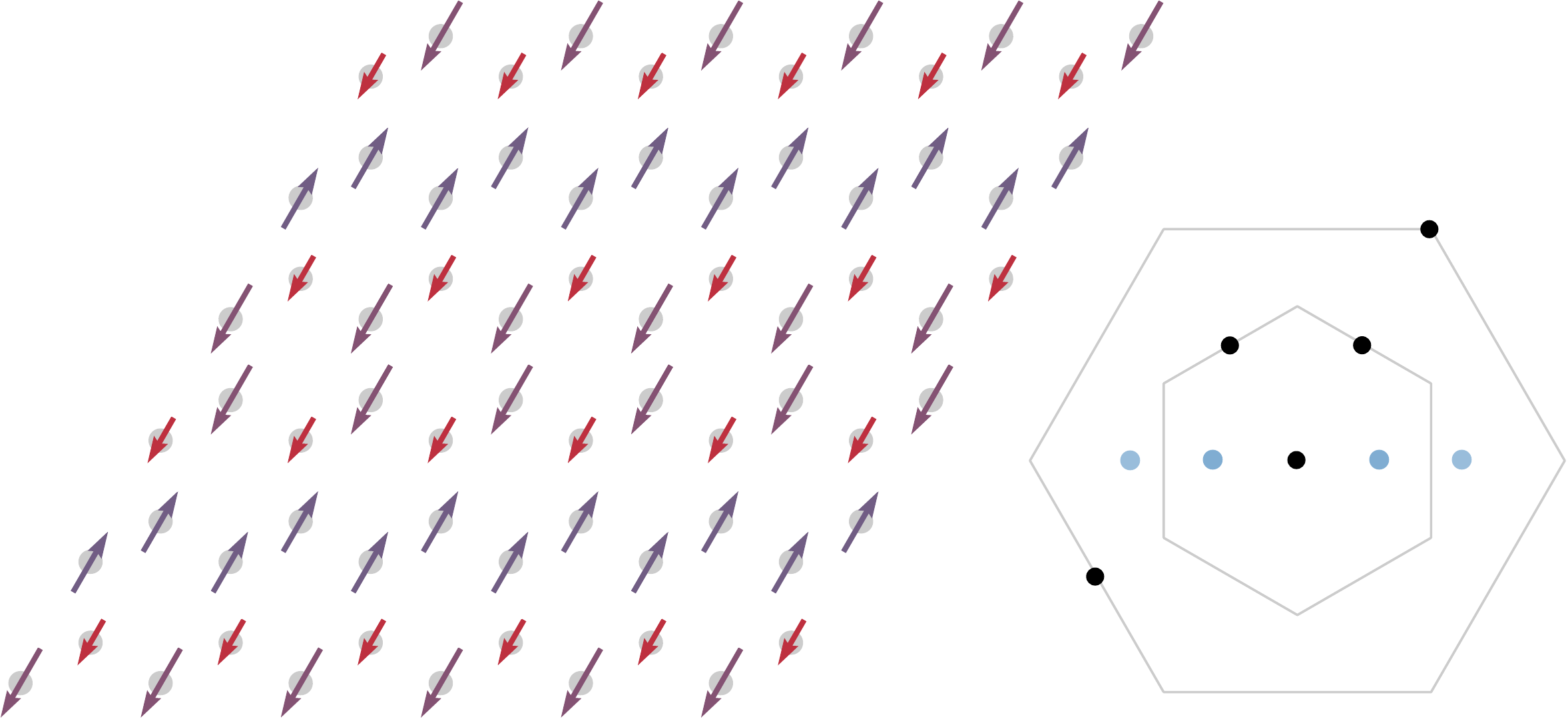}
    \put(10,35){(a)}
    \put(60,5){(b)}
    \put(93,30){$\Gamma'$}
    \put(66,7){$X$}
    \put(76,24){$Y$}
    \put(87,21){$M$}
    \put(84,15){$\Gamma$}
  \end{overpic}
\caption{
\label{fig:spins-2m3}(a) Spin configuration in the
ferrimagnetic $2M/3$ phase,
with the spins projected into the honeycomb plane. Deviation
out of the honeycomb plane is indicated by colour, with red being into
and blue being out of the plane. (b) Static structure factor for
this spin configuration, with intensity of colour denoting size of
$\vec{S}_Q \cdot \vec{S}_{-Q}$.
}
\end{figure}

If we are in the FK-zigzag regime in \nairo{} then we may be nearby in 
parameter space for \liiro{}, with a smaller negative or positive
value for $\Gamma'$. In the classical analysis this puts us in the
regime of phase I. At for the finite-size clusters used in these calculations
the ground state in this regime is given by a multiple-$Q$ state with a dominant
wave vector at $2M/3$, as mentioned in Sec. \ref{sec:classical}. This state has a 
tripled unit-cell and carries a finite ferrimagnetic moment of 
$\sim 0.254$ per site. The spin
configuration projected into the honeycomb plane is shown in 
Fig. \ref{fig:spins-2m3}. The spin-wave spectrum 
in this phase for $J=-3\meV$, $K=-25\meV$, $\Gamma=20\meV$ 
and $\Gamma'=0\meV$ is shown in Figs. \ref{fig:sw-inc}.
Besides a sizable spin-wave gap of order $\sim 4\meV$, we would like to
draw attention to the minima in the spin-wave spectrum near multiples
of $M/3$, in particular near $2M/3$. This is due to proximity
to a classical degeneracy when $J=\Gamma'=0$ and there are gapless
spin-waves.
This appears as a low-energy
dip in the integrated spectrum near $\sim 0.4 \AA^{-1}$. We note that
the magnitude of this wave-vector the inelastic features are in rough
agreement with recent reports of inelastic neutron scattering\footnote{S. K. Choi, \href{http://meetings.aps.org/link/BAPS.2014.MAR.W39.2}{APS March Meeting Talk, Denver (2014)}}
We note that these results hinge on the reliability
of the classical ground states. Given the small energy scales involved, quantum
mechanical effects may change some of the details of the state. 
Given the difference in trigonal distortion from \nairo{}, a state
related to this $2M/3$ phase may be relevant for \liiro{}.

\section{Conclusions}
\label{sec:conclusions}
In summary, we have analyzed a minimal nearest neighbour spin model for the honeycomb iridates
including the effects of trigonal distortion derived from
microscopic and symmetry arguments. Using classical simulated annealing
calculations and exact diagonalization of the full quantum mechanical model,
we mapped out the effects of this perturbation on the phase diagram. We 
identified two distinct zigzag phases: the AFK-zigzag and the FK-zigzag which
is stabilized by trigonal distortion. Based on the difficulty of finding an
AFK coupling while simultaneously tuning $\Gamma$ to be small, we argued that the FK-zigzag
with significant $\Gamma$ and small negative $\Gamma'$
is the likely candidate for \nairo{}. 
Looking at the dynamical structure factor within semi-classical spin-wave theory,
 we showed that this FK-zigzag phase can be made qualitatively consistent
with the experimental INS and RIXS data. We further showed that decreasing
the trigonal distortion in this parameter regime we find a family of multiple-$Q$ incommensurate
spiral phases. Given the smaller trigonal distortion in \liiro{}, we discussed
whether this phase I could be relevant for the magnetically ordered phase in this
material. Using a nearby commensurate phase with dominant wave-vector $2M/3$, we showed that 
this spiral phase is expected to have a gap with the lowest energy scattering 
occurring near wave-vector $2M/3$.

\emph{Note added:} After completion of this work, two preprints
appeared; Ref. \onlinecite{kimchi2014unified} discussing the nature of the 
ordered phase of \liiro{} and Ref. \onlinecite{sizyuk2014importance} discussing 
the origins of the zigzag phase in \nairo{}.

\acknowledgments

We would like to thank Eric Kin-Ho Lee and Yong-Baek Kim for useful discussions.
Computations were performed on the GPC supercomputer at the SciNet HPC 
Consortium.  SciNet is funded by: the Canada Foundation for Innovation under 
the auspices of Compute Canada, the Government of Ontario, Ontario Research 
Fund - Research Excellence; and the University of Toronto. This work was 
supported by the NSERC of Canada and the Centre for Quantum Materials 
at the University of Toronto.
HYK acknowledges the hospitality of Aspen Center for Physics (NSF Grant No. PHYS-1066293), 
where this work was finalized.
\bibliography{draft}

\appendix
\section{Strong-coupling expansion}
\label{app:general}
Since the atomic states of Ir are most easily presented using a quantization axis
that goes perpendicular to the honeycomb plane, it is natural to work in a pseudo-spin
with these quantization axes. The pseudo-spin model is then given by
\begin{eqnarray}
  H =\sum_{\avg{ij} \in \gamma} \Big[  &&
    J_1 S^z_i S^z_j + 
    \frac{J_2}{2}\left( S^+_iS^-_j +S^-_iS^+_j\right) \nonumber \\
&&    {J_3}\left( S^+_iS^+_je^{+i\phi_\gamma} +S^-_iS^-_j e^{-i\phi_\gamma}\right)+ \nonumber \\
&&    {J_4}\left(
 S^z_iS^+_je^{-i\phi_\gamma} +S^+_iS^z_je^{-i\phi_\gamma}\right)\nonumber \\
&& J_4 \left( S^z_iS^-_je^{+i\phi_\gamma} +S^-_iS^z_je^{+i\phi_\gamma}
\right)
 \Big], \\ \nonumber
\end{eqnarray}
where the exchanges $J_1$,$J_2$,$J_3$ and $J_4$ are related to $J$, $K$, $\Gamma$
and $\Gamma'$ through
\begin{subequations}
\begin{eqnarray}
  \label{eq:relations}
  J &=& \frac{1}{3}\left(J_1+2J_2-2J_3+2\sqrt{2}J_4\right), \\
  K &=& 2 \left(J_3 - \sqrt{2} J_4\right),\\
  \Gamma &=& \frac{1}{3}\left(J_1-J_2 + 4J_3 +2\sqrt{2} J_4\right),\\
  \Gamma' &=& \frac{1}{3}\left(J_1-J_2-2J_3-\sqrt{2}J_4\right).
\end{eqnarray}
\end{subequations}
To work out the strong-coupling expansion we first rotate the kinetic parts 
in Eq. \ref{eq:kinetic} into the $[111]$ quantization axes and carry out the perturbation
theory in this basis. Once complete, we use the relations above in Eq. \ref{eq:relations}
to find $J$,$K$,$\Gamma$ and $\Gamma'$ in the cubic axes. 
The full expressions for the exchanges discussed in Sec. \ref{sec:microscopics} are given by
\begin{widetext}
\begin{subequations}
\begin{eqnarray}
  \label{eq:exchanges}
  J &=& \frac{4}{27}\Bigg[
    \frac{
      6t_1(t_1+2t_3) - 9 t_4^2
    }{U-3J_H}
    +
    \frac{9 t_4^2 + 2( t_1-t_3)^2}{U-J_H}
    +
    \frac{
      (2t_1+t_3)^2
    }{U+2 J_H}
    +\\
&&\sqrt{2}\left(
\frac{3 (t_2 (4 t_3-5 t_4)+(7 t_1+t_3-4 t_4) t_4)}{3 J_H-U}
+\frac{4t_2(4 t_1-t_3)+t_4(11 t_1+13 t_3)+3t_4(4 t_4+5 t_2)}{J_H-U}
-\frac{4 (2 t_1+t_3) (t_2+2 t_4)}{2 J_H+U}
\right)\theta \Bigg],\nonumber
    \\
  K &=& \frac{8 J_H}{9}\left[
    \frac{(t_1-t_3)^2-3(t_2^2-t_4^2)}{(U-3J_H)(U-J_H)}+
    \frac{ \sqrt{2} \left(3  (t_2-t_4) t_4+ (t_3-t_1) (2 t_2+t_4)\right) }
    {(U-3J_H)(U-J_H)} 
  \theta\right],
    \\
  \Gamma &=& \frac{8 J_H}{9} \left[
    \frac{3t^2_4 + 2t_2(t_1-t_3)}{(U-3J_H)(U-J_H)}
-\frac{\sqrt{2}\left(t_2^2+(t_1-t_3)^2+(t_1-3 t_2-t_3) t_4+5 t_4^2\right)}
{(U-3J_H)(U-J_H)}\theta
\right],
\\
  \Gamma' &=& -\frac{8 J_H}{9} \left[
    \frac{t_4(t_1-t_3-3t_2)
    }{(U-3J_H)(U-J_H)}-
\frac{\sqrt{2}\left(2 t_1 (t_2+t_3)-t_1^2-(t_2+t_3)^2+ (3 t_2+t_3-t_1) t_4-4 \ (t_2^2+t_4^2)\right) }
{2(U-3J_H)(U-J_H)}\theta\right],
\end{eqnarray}
\end{subequations}
\end{widetext}
where we have expanded to leading order in trigonal distortion.
To recover the results 
of Ref. \onlinecite{rau2014generic} for the ideal octahedra,
simply take $\Delta/\lambda \rightarrow 0$.

\end{document}